\documentstyle[sprocl]{article}

\input{psfig}

\def\Journal#1#2#3#4{{#1} {\bf #2}, #3 (#4)}

\def\MPL{{\em Mod. Phys. Lett.} A}

\def\NPB{{\em Nucl. Phys.} B}

\def\PRD{{\em Phys. Rev.} D}

\def\al{\alpha}

\def\be{\begin{equation}}
\def\ee{\end{equation}}
\def\bea{\begin{eqnarray}}
\def\eea{\end{eqnarray}}
\begin{document}

\title{Gauge Symmetry Breakdown due to Dynamical Higgs Scalar
\footnote{Talk presented by T. Matsuki at the International Symposium on
QCD and Color Confinement held at RCNP, Osaka Univ, March, 2000.}}

\author{Takayuki MATSUKI}

\address{Tokyo Kasei University, 1-18-1 Kaga, Itabashi, Tokyo 173-8062, JAPAN
\\E-mail: matsuki@tokyo-kasei.ac.jp} 

\author{Masashi SHIOTANI}

\address{Faculty of Science and Technology, Kobe University, 1-1 Rokkohdai,
\\ Nada, Kobe 657, JAPAN\\E-mail: shiotani@tanashi.kek.jp}

\author{Richard HAYMAKER}

\address{Dept. of Physics and Astronomy,
Louisiana State University,
Baton Rouge,\\ LA 70803-4001, USA\\E-mail: haymaker@phys.lsu.edu}


\maketitle\abstracts{Assuming dynamical spontaneous breakdown of chiral
symmetry for massless gauge theory without scalar fields, we present a method
how to construct an effective action of the dynamical Nambu-Goldstone bosons
and elemetary fermions by using auxiliary fields. Here dynamical particles are
asssumed to be composed of elementary fermions. Various quantities including
decay constants are calculated from this effective action. This technique is
also applied to gauge symmetry breakdown, $SU(5)\rightarrow SU(4)$, to obtain
massive gauge fields.}

\section{Introduction}
\subsection{Massless $U(1)$ Gauge Theory}\label{U(1)}
Bound states composed of fermion-anti-fermion pairs can often be expressed
as bilocal fields. Those fields are so far treated classically using the
Bethe-Salpeter equation. Or only its vacuum expectation vaule is caluculated
since it is also a c-number. Here in this report we present a method~\cite{ms}
how to treat those bilocal fields as local ones so that one can use Feynman
diagrams to calculate physical quantities. That is, we propose a method by
which one can treat bilocal fields as quantum ones at least in the first
order of approximation.

We use massless $U(1)$ gauge theory to descibe our idea as the simplest
example. Starting from the following Lagrangian,
\begin{equation}
  {\cal L}_0=-\frac{1}{4}\left(F^{\mu\,\nu}\right)^2-\frac{1}{2\alpha}\left(
  \partial_\mu A^\mu\right)^2+\bar\psi\;i D{\kern -7pt /}\;\psi,\label{L0}
\end{equation}
where $D_\mu=\partial_\mu-igA_\mu$ and $\al$ is a gauge parameter. We just
describe how to obtain the effective Lagrangian below. First functionally
integrate patition function using Eq.(\ref{L0}) over gauge fields. Then
the resultant four-fermi interactions are cancelled by introducing bilinear
terms of auxiliary fields~\cite{hm}. Finally we are left with the effective
Lagrangian including fermions as well as auxiliary fields which are decomposed
into local fields as
\begin{eqnarray}
  S_{\rm eff}&=&\int\;d^4x\;\bar\psi(x)\;i \partial{\kern -6pt /}\;\psi(x)
  -\int\;d^4x\,d^4y\;\bar\psi(x)\Big[\phi(x,y)+i\gamma_5\pi(x,y)\Big]\psi(y)
  \nonumber \\
  &+&\frac{1}{2}\int\;d^4x\,d^4y\;\Big[\phi(x,y)D^{-1}(x-y)\phi(y,x)+
  \pi(x,y)D^{-1}(x-y)\pi(y,x)\Big]+\ldots, \label{Yukawa}
\end{eqnarray}
with
\begin{eqnarray}
  D(x-y)&=&\frac{g^2}{4}\,g^{\mu\,\nu}D_{\mu\,\nu}(x-y)=
  -\frac{\lambda}{\left(x-y\right)^2},\\
  \phi(x,y)&=&\frac{\phi_0(x-y)}{f}\left[f+\sigma\left(\frac{x+y}{2}
  \right)\right], \label{phi} \\
  \pi(x,y)&=&\frac{\phi_0(x-y)}{f}\;\varphi\left(\frac{x+y}{2}\right).
  \label{pi}
\end{eqnarray}
and a gauge boson propagator $D_{\mu\,\nu}$, $D^{-1}=1/D$, and
$\lambda=(3+\al)g^2/(16\pi^2)$. Eqs.~(\ref{phi}, \ref{pi}) are the key point of
our paper. This decomposition of these fields, $\phi(x,y)$ and $\pi(x,y)$ can
be interpreted as a decomposition of internal degrees of freedom depending on
$x-y$, i.e., $\phi_0(x-y)$, and a total degree of freedom depending on
$(x+y)/2$, i.e., $\sigma((x+y)/2)$ and $\varphi((x+y)/2)$ which are treated as
local fields. Translational invariant quntity $\phi_0(x-y)$ is regarded as a
vacuum expectation value of a bilocal field $\phi(x,y)$. This intuitive
interpretation can also be supported by chekcing that mass of the Nambu-Golston
(NG) boson vanishes by calculating only bilinear terms in $\varphi$ which
consist of tree term and fermion one-loop diagram from Eq.(\ref{Yukawa}).
The coefficients of these two bilocal fields become the same, $\phi_0(x-y)$, can
be derived by this vanishing NG boson mass. This $\phi_0(x-y)$ satisfies the
gap equation, namely it gives a fermion mass $\Sigma(q)$ as
\begin{equation}
  \partial_q^2\Sigma(q)+\frac{4i\lambda\Sigma(q)}{q^2-\Sigma^2(q)}=0,
  \label{gap1}
\end{equation}
with $\Sigma(q)=\int\,d^4r/(2\pi)^4\,\phi_0(r)\exp(-iqr)$.
\subsection{Decay Constant in $U(1)$ Gauge Theory}\label{decay}
Using this result, the physical quntity to be calculated first is a decay
constant in the $U(1)$ case defined by
\begin{equation}
  \left<0\left|\bar\psi(0)\;\gamma_\mu\gamma_5\psi(0)\right|\varphi(q)\right>=
  -2i\,f\,q_\mu. \label{Dconst}
\end{equation}
The Bethe-Salpeter equation is used to describe the left hand side of
Eq.~(\ref{Dconst}).~\cite{kugo} Our method tells us to calculate just one
fermion loop which connects an axial vector vertex and the NG boson,
$\varphi(q)$. Our method reproduces the so-called Pagels-Stokar~\cite{ps}
formula for a decay constant $f$,
\begin{equation}
  f^2=\frac{1}{2\left(2\pi\right)^2}\,\int\,xdx\,
  \frac{\Sigma(x)\Bigl[\Sigma(x)-x\Sigma'(x)/2\Bigr]}
  {\Bigl[x+\Sigma^2(x)\Bigr]^2},
\end{equation}
where momentum space is converted into Eucledean space, i.e., $x=-p^2$.
\section{Dynamical Breakdown of Massless $SU(5)$ Gauge Theory}
The next application of our method is to calculate gauge boson masses when
gauge symmetry dynamically breaks down. We study the case in which massless
$SU(5)$ gauge theory dynamically breaks down to $SU(4)$ which was predicted
as the most favorable breaking pattern due to a tumbling scenario.~\cite{rds}
However nobody has ever shown how to calculate gauge boson masses when symmetry
breaks down. Peole have just predicted that masses may be obtained by applying
the Pagels-Stokar formula and multiplying a gauge coupling constant since there
is no way to calculate those qunatities so far. We will show our method can do
this job.

The difference between this section and Subsections \ref{U(1)} and \ref{decay}
is that whether there is a real gauge field which couples to a fermion axial
vector vertex or not. In the former section we do not have such a coupling,
while in this gauge theory we do.

Now we start from the following Lagrangian
\bea
  {\cal L}_0&=&-\frac{1}{4}\left(F^A_{\mu\;\nu}\right)^2
  +\;i\left(\overline\psi\right)^i\left[{\partial\kern-0.6em /}\;\delta_i{}^j-
  ig{A^A\kern-1.2em /}{\kern+0.7em}\left(T^A\right)_i{}^j\right]\psi_j
  \nonumber \\
  &\quad& +\frac{i}{2}{\left(\overline\chi\right)}{~}^{i\;j}\left[
  {\partial\kern-0.6em /}{\kern+0.4em}\delta_j{}^k-2ig{A^A\kern-1.2em /}
  {\kern+0.5em}\left(T^A\right)_j{}^k\right]\chi_{k\;i},\label{L0}
\eea
where $A=1\sim24$, $(\psi^c)^i\sim\underline{5}^*$ and $\chi_{i\,j}\sim
\underline{10}$ left
handed fermions are introduced. Our breaking pattern tells us that
5-dimensional $\phi_i\sim\underline{10}\times \underline{10}$ complex scalar
bound state plays a role of Higgs scalar, which should give mass to
\underline{1} and $\underline{4}/\underline{4}^*$ representations out of
$\underline{24}=\underline{15}+\underline{4}+\underline{4}^*+\underline{1}$
representations under $SU(4)$. \underline{15} corresponds to massless $SU(4)$
gauge fields which are functionally integrated out and give four-fermi terms
that are cancelled by bilinear terms in the bilocal auxiliary fields.

Corresponding to Eqs.~(\ref{phi}, \ref{pi}), we have
\bea
  \phi_i(x,y) &=& \left\{\frac{\phi_0(x-y)}{v}\left[v+
  \sigma\left(\frac{x+y}{2}\right)\right]\exp\left[\frac{i\pi^\alpha
  ((x+y)/2) T^\alpha}{v}\right]\eta\right\}_i
  \\
  &=& \frac{\phi_0(x-y)}{v}\left[v+
  \sigma\left(\frac{x+y}{2}\right)\right]\delta_i{}^5
  +i\frac{\phi_0(x-y)}{v}\pi^\alpha\left(\frac{x+y}{2}\right)
  \left(T^\alpha\right)_i{}^5+\ldots,
  \nonumber \\
  \eta_i &\equiv& \delta_i{}^5.
\eea
Using this expression, we can calculate gauge boson masses, \underline{1}
and \underline{4}. However these are related to \underline{6} fermion mass
$\Sigma(q)=-2\int\,d^4r/(2\pi)^4 \phi_0(r)\exp(-iqr)$ which is the only fermion
mass available. Fermion mass is only obtained after $SU(5)$ breaks down to
$SU(4)$, which means mass ratio between \underline{1} and \underline{4} gauge
bosons is not given by that as expected from an elementary Higgs. In a
sense, gauge boson masses are available after gauge symmetry breaks down due to
fermion chiral condensation occurs. All these details will be given in a
separate paper.~\cite{hms}

\section*{Acknowledgments}
M. S. acknowledges Prof. T. Morii for his continuous encouragements during the
course of this study. T. M. and M. S. would also like to thank the theory
group of KEK at Tanashi where part of this work has been done.


\section*{References}
\begin{thebibliography}{99}
\bibitem{ms}T. Matsuki and M. Shiotani, \Journal{\MPL}{10}{709}{2000}.

\bibitem{hm}R.W.~Haymaker, T.~Matsuki, and F.~Cooper, 
\Journal{\PRD}{35}{2567}{1987}; C.D.~Roberts and R. T.~Cahill, \Journal
{\PRD}{33}{1755}{1986}; T.~Morozumi and H.~So, {\em Prog. Theor. Phys.} 77, 
1434 (1987).

\bibitem{kugo}
  K. Aoki, M. Bando, T. Kugo, M.G. Mitchard, H. Nakatani, Prog. Theor. Phys.
 {\bf 84}, 683 (1990);

\bibitem{ps}
  R. Jackiw and K. Johnson, \Journal{\PRD}{8}{2386}{1973};
  H. Pagels and S. Stokar, \Journal{\PRD}{20}{2947}{1979}.

\bibitem{rds}S. Raby, S. Dimopoulos, and L. Susskind, 
\Journal{\NPB}{169}{373}{1980}.

\bibitem{hms}H.W. Haymaker, T. Matsuki, and M. Shiotani, preprint in
preparation.

\end{thebibliography}
\end{document}